\documentclass[preprint]{aastex}

\usepackage{graphicx}
\usepackage{longtable}
\usepackage{lscape}
\usepackage{natbib}

\usepackage[usenames,dvipsnames,svgnames,table]{xcolor}

\shorttitle{The magnetic sensitivity of the Mg {\sc ii} $k$ line}
\shortauthors{Alsina Ballester, Belluzzi \& Trujillo Bueno}

\begin{document}

\title{The magnetic sensitivity of the Mg\,{\sc ii}~k line to\\
the joint action of Hanle, Zeeman and magneto-optical effects}

\author{E. Alsina Ballester\altaffilmark{1,2}, 
L. Belluzzi\altaffilmark{3,4}, and 
J. Trujillo Bueno\altaffilmark{1,2,5}} 

\altaffiltext{1}{Instituto de Astrof\'{\i}sica de Canarias, E-38205  
La Laguna, Tenerife, Spain}
\altaffiltext{2}{Departamento de Astrof\'{\i}sica, 
Universidad de La Laguna, E-38206 La Laguna, Tenerife, Spain}
\altaffiltext{3}{Istituto Ricerche Solari Locarno, CH-6605 Locarno Monti, 
Switzerland}
\altaffiltext{4}{Kiepenheuer-Institut f\"ur Sonnenphysik D-79104 Freiburg, 
Germany}
\altaffiltext{5}{Consejo Superior de Investigaciones Cient\'{\i}ficas, Spain}

\email{ealsina@iac.es, belluzzi@irsol.ch, jtb@iac.es}

\begin{abstract}
We highlight the main results of a radiative transfer investigation on the 
magnetic sensitivity of the solar Mg~{\sc ii} $k$ resonance line at 2795.5 \AA, 
accounting for the joint action of the Hanle and Zeeman effects 
as well as partial frequency redistribution (PRD) phenomena. 
We confirm that at the line center, the linear polarization signals
produced by scattering processes are measurable, and that they are sensitive, 
via the Hanle effect, to magnetic fields with strengths between 5 and 50~G, 
approximately. We also show that the Zeeman effect produces conspicuous circular 
polarization signals, especially for longitudinal fields stronger than 50~G, 
which can be used to estimate the magnetization of the solar 
chromosphere via the familiar magnetograph formula. 
The most novel result is that magneto-optical effects produce, in the wings of the
line, a decrease of the $Q/I$ scattering polarization pattern and the appearance of $U/I$ signals (i.e., a rotation
of the plane of linear polarization). This sensitivity of the $Q/I$ and $U/I$ wing signals to both weak (${\sim}\,5$ G) and stronger
magnetic fields expands the scientific interest of the Mg {\sc ii} $k$ line for probing the chromosphere in quiet and active regions of the Sun.

\end{abstract}


%%%%%%%%%%%%%%%%%%%%%%%%%%%%%%%%%%%%%%%%%%%%%%%%%%%%%%%%%%%%
\section{Introduction}
The 
inference of the magnetic field in the upper chromosphere and 
transition region of the Sun is a very important challenge of modern astrophysics. 
In such external atmospheric 
layers, the temperature of the solar plasma 
ranges from $10^4$~K to $10^5$~K, so that its primary emission lies at 
wavelengths shorter than 3000~{\AA}. 
Since the magnetic field leaves its fingerprints on the spectral line 
polarization \citep[e.g.,][]{Stenflo94,LL04}, in order to probe the magnetic 
properties of the 
outer solar atmosphere we need to measure 
and interpret the four Stokes profiles ($I, Q, U, V$) of strong ultraviolet 
(UV) spectral lines \citep[e.g., the review by][and more references therein]{Trujillo14}.

Recent investigations pointed out that the hydrogen Ly-$\alpha$ line of the solar disk 
radiation should be linearly polarized, and that via the Hanle effect the line-center polarization 
is sensitive to the presence of magnetic fields, with strengths between 10 and 100 
gauss, in the corrugated boundary that delineates  
the chromosphere-corona transition region \citep[see][]{Trujillo+11,Trujillo+12, Belluzzi+12,
Stepan+12}. Such theoretical advances motivated the development of the Chromospheric 
Lyman-Alpha Spectro-Polarimeter (CLASP), a suborbital rocket experiment that has provided the
 first successful measurement of the $Q/I$ and $U/I$ profiles of the Ly-$\alpha$ line at 1216~\AA\ in
 relatively quiet regions of the solar disk \citep[][]{Kano+16}.  

The success of CLASP and the results of a recent theoretical investigation 
\citep[see][]{BelluzziTrujillo12} have led the CLASP international team 
to propose a second flight aimed at observing the four Stokes profiles 
across the Mg~{\sc ii} $h$ \& $k$ lines around 2800~\AA. 
In their theoretical work, \cite{BelluzziTrujillo12} investigated the linear 
polarization signals of these lines due to scattering processes and the circular polarization due to 
the Zeeman effect. In particular, they demonstrated that the joint action of partial frequency
redistribution (PRD) and quantum interference between the upper 
$J$-levels of the $h$ \& $k$ lines (hereafter, $J$-state interference) produces 
a complex scattering polarization $Q/I$ pattern showing extended wings with  
sizable amplitudes, and a positive peak at the center of the $k$ line 
surrounded by two negative peaks (see the solid curve of Fig. 1). 
They also pointed out that the two-level atom approximation is 
suitable for modeling the center and near-wings of the Mg {\sc ii} $k$ line, including the $Q/I$ negative 
peaks surrounding the positive line-center one (compare the solid and dashed 
curves of Fig. 1). 

Within the framework of a rigorous quantum theory of spectral line 
polarization \citep{Bommier97}, we have developed a PRD two-level atom 
radiative transfer (RT) code taking into account collisional and radiative transitions and the joint action 
of scattering processes and the Hanle and Zeeman effects produced 
by arbitrary magnetic fields \citep[see][]{Alsina+16}. 
In the illustrative calculations presented in that paper, we pointed out that 
in strong resonance lines the magneto-optical terms $\rho_V\,U$ and $\rho_V\,Q$ 
of the transfer equations for Stokes $Q$ and $U$, respectively, can produce an 
interesting magnetic sensitivity in the wings of the $Q/I$ and $U/I$ 
profiles. The aim of the present paper is to show and discuss the results we have 
obtained for the 
four Stokes parameters of the Mg~{\sc ii} $k$ line, 
highlighting the interesting magnetic sensitivity that results from the joint 
action of Hanle, Zeeman and magneto-optical effects. 

Since we are using the two-level atom approximation, which by definition cannot 
account for the effects of $J$-state interference, our results apply mainly to 
the center and near-wings of the Mg~{\sc ii} $k$ line (i.e., to the spectral 
region between 2795~\AA\ and 2796~\AA), which encode information on the 
magnetism of the solar chromosphere. The next step of our work will be to 
consider the case of a two-term atom, by generalizing the PRD with $J$-state interference 
approach that \cite{BelluzziTrujillo14} developed for the unmagnetized reference case.

%%%%%%%%%%%%%%%%%%%%%%%%%%%%%%%%%%%%%%%%%%%%%%%%%%%%%%%%%%%%
\section{Formulation of the problem} 
The upper levels of the Mg~{\sc ii} $h$ and $k$ lines have total angular 
momenta $J=1/2$ and $J=3/2$, respectively, while the lower level, which is 
common to the two spectral lines, has $J=1/2$.
Both resonance lines are sensitive to the Zeeman effect but, observing that 
atomic levels with $J=1/2$ cannot carry atomic alignment, only the Mg~{\sc ii} 
$k$ line is sensitive to the Hanle effect. 
The critical magnetic field for the onset of the Hanle effect in the 
Mg~{\sc ii} $k$ line is $B_{\rm H}=22$~G, which means that in the presence of 
magnetic fields with strengths $B > 0.2 B_H \approx 5~G$ we can expect a 
significant modification of the {\em line-center} amplitudes of the $Q/I$ and 
$U/I$ profiles with respect to the zero-field reference case. 

%%%%%%%%
\begin{figure}[t]
\begin{center}
\includegraphics[scale=0.45]{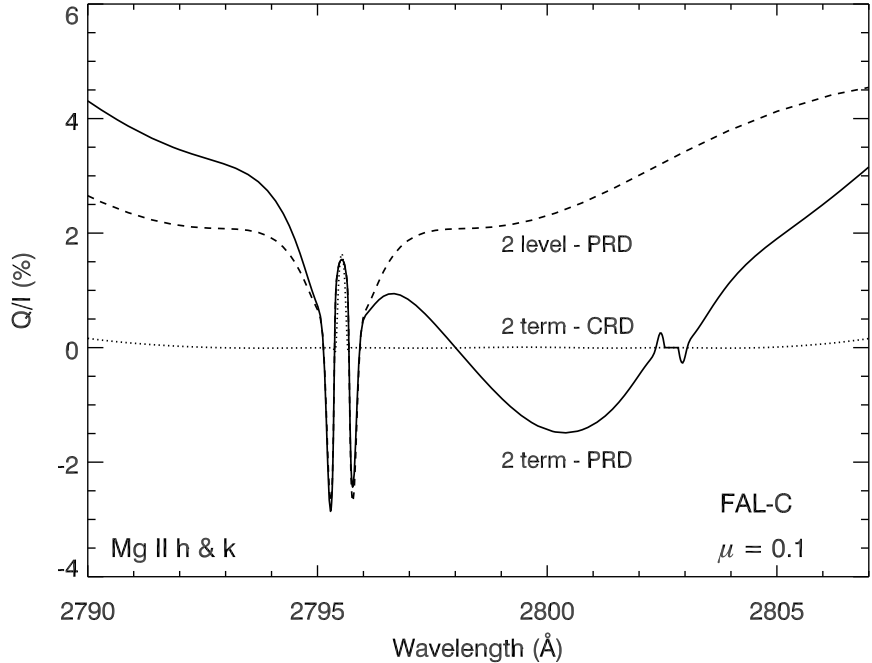}
\caption{The scattering polarization $Q/I$ pattern of the Mg~{\sc ii} $h$ \& 
$k$ lines calculated in the semi-empirical model C of \cite{Fontenla+93} 
(hereafter; FAL-C) assuming CRD (dotted curve) and taking into account PRD 
effects (solid curve), in both cases including the effects of $J$-state 
interference. 
The dashed curve indicates the two-level atom PRD solution for the Mg~{\sc ii} 
$k$ line; note that it is a reasonable approximation for modeling the positive 
line-center peak and the two negative peaks located in the near wings of the 
$Q/I$ profile. 
The reference direction for positive Stokes $Q$ is the parallel to the 
nearest limb.}
\label{Figure-1}
\end{center}
\end{figure} 
%%%%%%%% 

As mentioned in \S1, 
the spectral region of the Mg~{\sc ii} $k$ line  
 between 2795~{\AA} and 2796~{\AA} is of great scientific interest. 
 In this region, 
 the scattering polarization $Q/I$ pattern  
shows a positive line-center peak surrounded by two negative peaks (see Fig.~1). 
The line-center peak is sensitive to the presence 
of magnetic fields via the Hanle effect, but what about the two 
negative peaks?
In order to investigate this issue, we relax the commonly-used weak-field 
approximation (i.e., the assumption that outside active regions the linear 
polarization is fully dominated by scattering processes, and the Zeeman 
splitting can be neglected in the absorption and emission profiles), and we 
therefore consider the joint action of 
 the Hanle and Zeeman effects. 
It can be easily shown that if the lower level is unpolarized (as it is 
assumed in this work), and the Zeeman splittings are neglected, the absorption 
coefficient for the intensity, $\eta_I$, is the only non-zero element of the 
propagation matrix \citep[see][]{LL04}, and the RT equations for the Stokes $Q$ 
and $U$ parameters read:
\begin{equation}
\frac{{\rm d}Q}{{\rm d}s} \, = \, \epsilon_Q \, - \, \eta_I \, Q \, , \\
\label{QRTE1}
\end{equation}

\begin{equation}
\frac{{\rm d}U}{{\rm d}s} \, = \, \epsilon_U \, - \, \eta_I \, U \, ,
\label{URTE1}
\end{equation}
with $s$ the geometrical distance along the ray, and $\epsilon_X$ 
the emissivity for the Stokes parameter $X$ (with $X=Q, U$).
However, if the Zeeman splitting of the line's atomic levels is taken into 
account, then the absorption coefficients $\eta_Q$, $\eta_U$ and 
$\eta_V$, as well as the magneto-optical terms $\rho_Q$, $\rho_U$ and $\rho_V$ 
are in general non-zero, and the transfer equations for the Stokes $Q$ 
and $U$ parameters are

\begin{equation}
\frac{{\rm d}Q}{{\rm d}s} \, = \, \epsilon_Q \, - \, \eta_Q \, I \, - \, 
\rho_V \, U \, + \, \rho_U \, V  \, - \, \, \eta_I \, Q \, \, 
{\approx} \, \, [ \epsilon_Q \, - \, \rho_V \, U ] \, 
- \, \, \eta_I \, Q \, ,
\label{QRTE2}
\end{equation}

\begin{equation}
\frac{{\rm d}U}{{\rm d}s} \, = \,  \epsilon_U \, - \, \eta_U \, I \, + \, 
\rho_V \, Q \, - \, \rho_Q \, V  \, - \, \eta_I \, U \, \, 
{\approx} \, \, [ \epsilon_U \, + \, \rho_V \, Q ] \, 
- \, \eta_I \, U \, .
\label{URTE2}
\end{equation}

%%%%%%%%%
\begin{figure}[t]
\begin{center}
\includegraphics[scale=0.33]{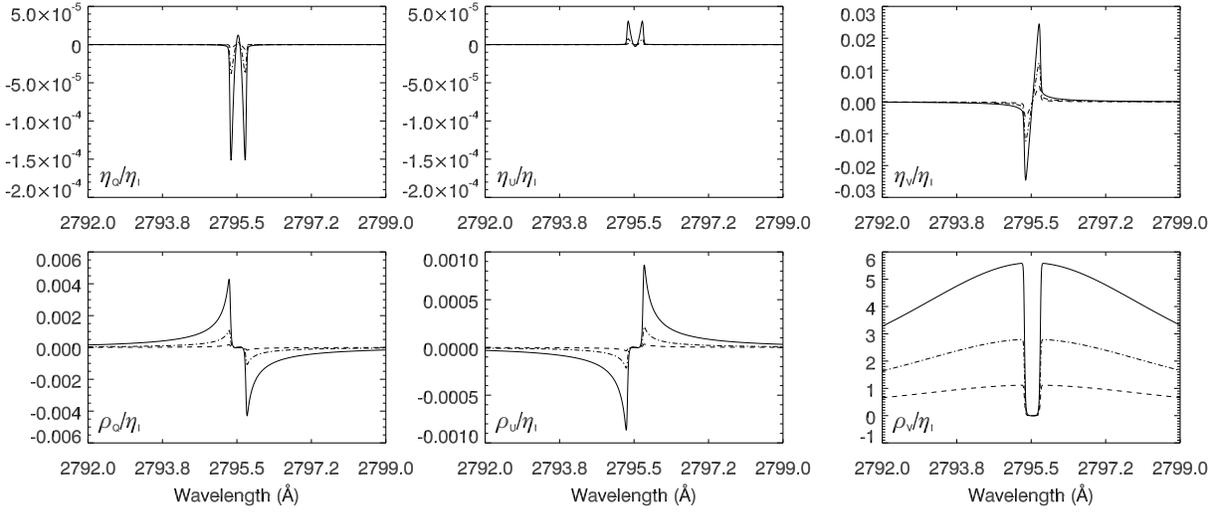}
\caption{Wavelength variation of  
$\eta_X/\eta_I$ and $\rho_X/\eta_I$ (with $X$ equal to $Q$, $U$ or $V$) in the 
FAL-C model atmosphere (at a height $h=1860$ km).
The magnetic field is horizontal with azimuth 
$\chi_B=45^{\circ}$ (see text).  
The considered magnetic strengths are 20 G (dashed 
curves), 50 G (dashed-dotted) and 100 G (solid).}
\label{Figure-2}
\end{center}
\end{figure} 
%%%%%%%% 

Fig.~2 shows the wavelength variation of $\eta_X/\eta_I$ and of $\rho_X/\eta_I$ (with 
$X$ equal to $Q$, $U$ or $V$) at a given chromospheric height in the FAL-C 
atmospheric model, where we have imposed a constant 
horizontal magnetic field with azimuth $\chi_B=45^{\circ}$ (measured counter-clockwise 
from the plane defined by the vertical and the line-of-sight). All the above-mentioned ratios 
are very small (or zero) at the line center, while they show 
either symmetric or antisymmetric peaks just outside the line-core region.
As expected, all such ratios increase with the magnetic strength. 
For magnetic fields weaker than about 100~G, the amplitude of the peaks is, 
however, very small, with the clear exception\footnote{We note that in the limit of weak fields 
the $\eta_V$ and $\rho_V$ quantities scale with the ratio, $\cal {R}$, between the Zeeman splitting and the line's
 width, while $\eta_X$ and $\rho_X$, with $X=\{Q, U \}$, scale with ${\cal {R}}^2$.}
 of $\eta_V/\eta_I$ and $\rho_V/\eta_I$. 
Interestingly, all such ratios rapidly tend to zero towards the blue and red wings of 
the Mg~{\sc ii} $k$ line, with the notable exception of $\rho_V/\eta_I$. 
The reason why in the line wings $\rho_V$, which couples Stokes $Q$ and $U$ 
(see Eqs.~3 and 4), is large compared to the absorption coefficient $\eta_I$ 
is because of the broad wings of the Faraday-Voigt profiles that appear in 
its expression \citep[see][]{LL04}. 
Note that the wing (e.g., $\lambda\,{<}\,2795$ \AA\ and $\lambda\,{>}\,2796$ \AA)
and near-wing (e.g., $\lambda\,{=}\,2795$ \AA\ and $\lambda\,{=}\,2796$ \AA) values of 
$\rho_V/\eta_I$ are significant already for magnetic fields as weak as 10 G 
(i.e., of the order of the Hanle critical field for the Mg {\sc ii} $k$ line). 
 For sufficiently weak magnetic fields (e.g., of the order of 100~G), 
the dominant terms of the RT equations for $Q$ and $U$ are therefore as indicated by the
approximate equalities in Eqs. (3) and (4).

Although $\rho_V/\eta_I$ is negligible in the line core region (see Fig.~2), in the wings 
the terms $\rho_V U$ and $\rho_V Q$ must, in general, be considered.
Such terms can only be neglected if at the wing and near-wing wavelengths 
Stokes $Q$ and $U$ are negligible, as it happens when the complete frequency 
redistribution (CRD) approximation is used (see the dotted curve of Fig.~1). 
However, in strong resonance lines, like the Mg~{\sc ii} $h$ \& $k$ lines, the 
effects of PRD and $J$-state interference produce complex scattering 
polarization profiles with extended wings (see the solid curve of Fig.~1). 
As soon as there is a magnetic field, even as weak as 5~G, the $\rho_V$ 
coefficient, which couples Stokes $Q$ and $U$, becomes significant and it is responsible for 
 the appearance of $U/I$ wing signals and 
a magnetic sensitivity of the wing and near-wing values of the $Q/I$ and $U/I$ profiles.

%%%%%%%%%%%%%%%%%%%%%%%%%%%%%%%%%%%%%%%%%%%%%%%%%%%%%%%%%%%%
\section{Results}

For the scientific 
 purpose of this paper, it suffices to discuss the results of our   
RT calculations in the FAL-C 
atmospheric model, 
in the presence of magnetic fields of different 
intensity and orientation. We show examples of emergent Stokes 
profiles of the Mg {\sc ii} $k$ line radiation, calculated for two 
different line-of-sights (LOS). Indicating with $\mu$ the cosine of the 
heliocentric angle, we consider the LOS corresponding to $\mu=0.1$ (close to 
the limb geometry) and $\mu=1$ (disk center geometry). 

%%%%%%%%%%%%%%%%%%%%%%%%%%%%%%%%%%%%%%%%%%%%%%%%%%%%%%%%%%%%
\subsection{Close to the limb geometry (LOS with $\mu=0.1$)}

%%%%%%%%
\begin{figure}[t]
\begin{center}
\includegraphics[scale=0.28]{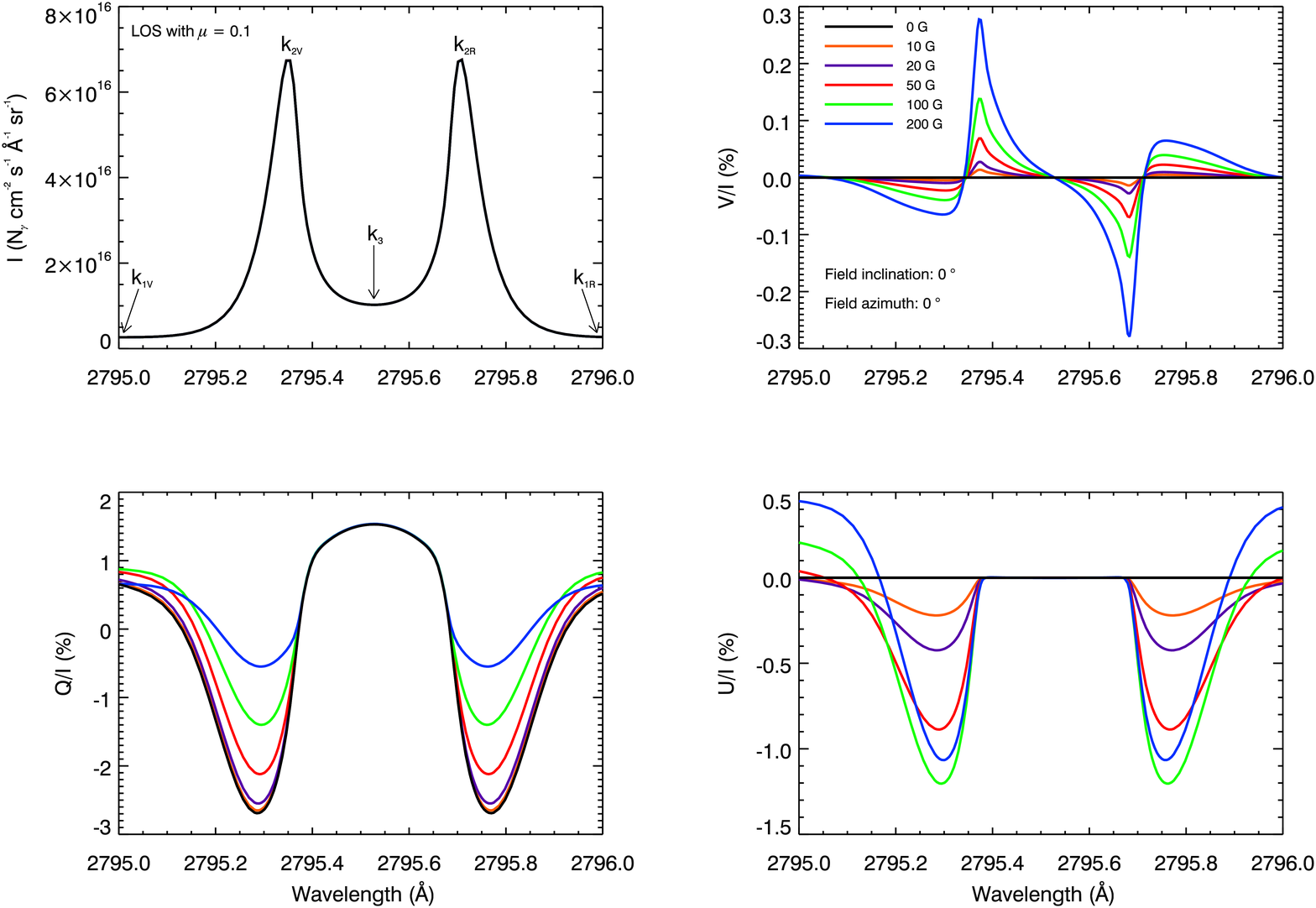}
\caption{Stokes profiles of the Mg {\sc ii} $k$ line calculated in the 
FAL-C semi-empirical model, in the absence (black curves) and in the 
presence (colored curves) of a {\em vertical} magnetic field for a LOS
with $\mu = 0.1$.
The reference direction for positive Stokes $Q$ is the parallel to the 
nearest limb. 
}
\label{Figure-3}
\end{center}
\end{figure} 
%%%%%%%%    

Figure 3 shows the emergent Stokes profiles in the absence  
and in the presence of a vertical magnetic field of increasing strength, from 
$B=0$ G till $B=200$ G. 
The black curves show the zero-field reference case, with a $Q/I$ pattern 
characterized by a central positive peak surrounded by two negative peaks, 
and $U/I=V/I=0$. 
As expected, in the presence of a vertical magnetic field there is no 
modification of the $Q/I$ and $U/I$ line-center values, because the applied 
vertical field cannot produce any Hanle effect (we recall that the 
Hanle effect only operates in the center of the spectral lines). 
Since for a LOS with $\mu=0.1$ the applied vertical field has a 
longitudinal component, there is a non-zero circular polarization $V/I$ pattern 
with positive and negative lobes, whose amplitudes increase with the 
magnetic strength. 
The most interesting feature is the very significant magnetic 
sensitivity seen in the wings of the $Q/I$ and $U/I$ profiles, which occurs 
already for magnetic fields as weak as 10~G. 
For the vertical field case, the wings of $Q/I$ are increasingly depolarized as 
the magnetic strength increases, while between zero and 100 G the wing values of $U/I$ increase. 
As advanced in \S2, the reported magnetic sensitivity of the $Q/I$ and $U/I$ 
wing signals is due to the $\rho_V\,U$ and $\rho_V\,Q$ terms of Eqs.~(3) and 
(4), respectively. Such magneto-optical terms have an important impact on the wings of the $Q/I$ 
and $U/I$ profiles of strong resonance lines like Mg {\sc ii} $k$, because 
outside the line center $\rho_V/\eta_I$ is very significant, already for weak 
magnetic fields (see Fig.~2), and because in strong resonance lines PRD 
effects produce broad scattering polarization profiles 
with sizable wing amplitudes \citep{BelluzziTrujillo12}. 
As a matter of fact, if the same RT calculations are 
performed neglecting the $\rho_V$ terms, 
$Q/I$ shows no magnetic sensitivity, and no $U/I$ signal 
is obtained.
%%%%%%%%
\begin{figure}[t]
\begin{center}
\includegraphics[scale=0.28]{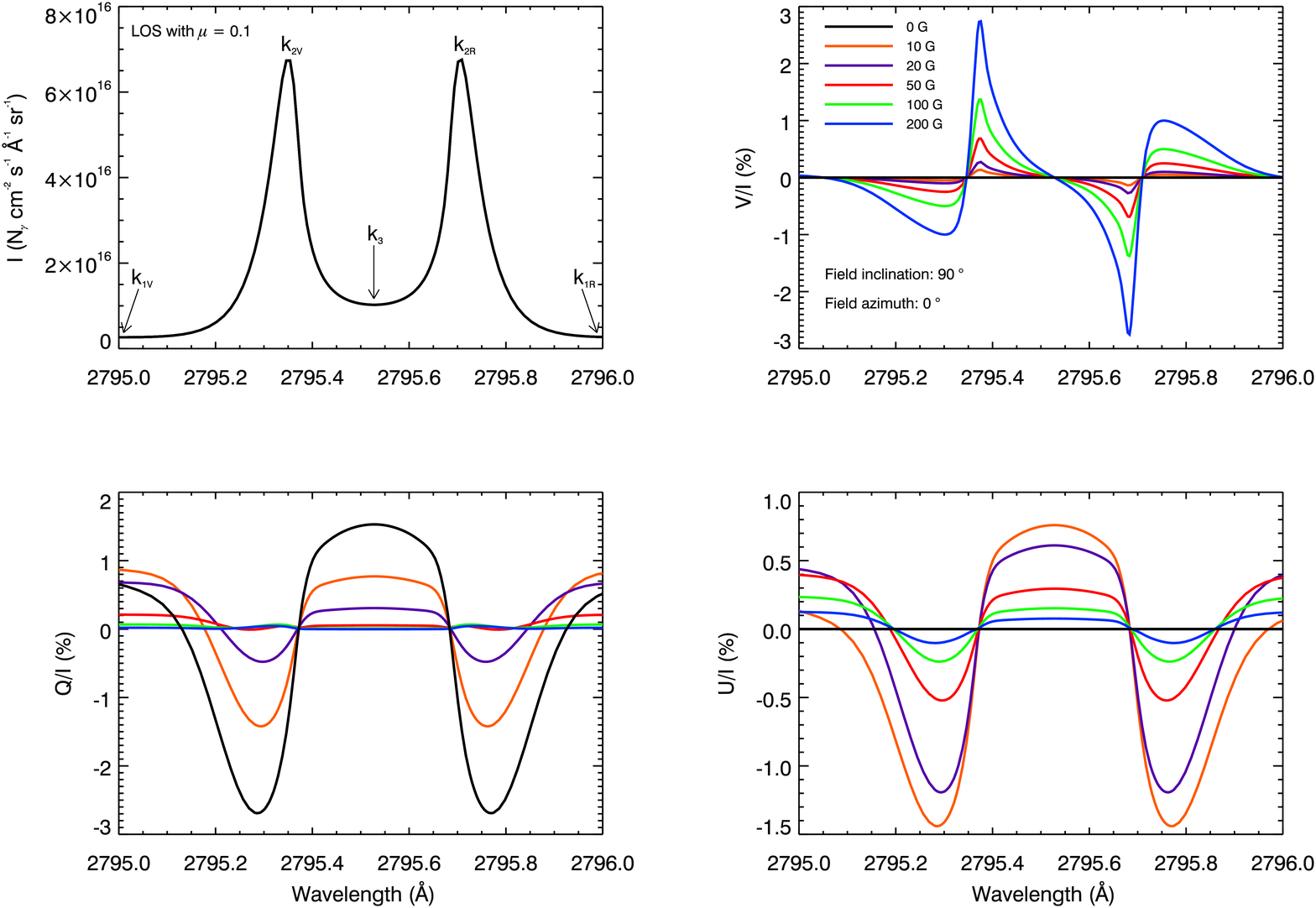} 
\caption{Same as Fig~3, but for a horizontal magnetic field 
with azimuth $\chi_B=0^{\circ}$. 
The reference direction for positive Stokes $Q$ is the parallel to the nearest limb.}
\label{Figure-4}
\end{center}
\end{figure} 
%%%%%%%%

We consider now the case of a magnetic field inclined with respect to the local 
vertical direction, so as to see the Hanle effect in action. 
Figure 4 shows the emergent Stokes profiles of the Mg {\sc ii} $k$ line 
calculated in the absence (black curves) and in the presence of a 
horizontal magnetic field with azimuth $\chi_B=0^{\circ}$. 
As expected, the polarization amplitudes at the center of the $Q/I$ and 
$U/I$ profiles are now modified by the Hanle effect. 
As the intensity of the magnetic field increases the $Q/I$ line-center 
amplitude decreases, while the $U/I$ line-center signal (which is zero in the 
absence of the magnetic field) first increases till a maximum ${\sim}1\%$ 
for $B\,{\approx}\,10$ G, and then decreases till reaching 0.1\% for 
$B=200$~G. Fig.~4 shows the above-mentioned very 
interesting magnetic sensitivity of the $Q/I$ and $U/I$ profiles in the
line wings, due to the $\rho_V$ magneto-optical term. Note that  
the wings of the $Q/I$ 
profile are also increasingly depolarized as the intensity of the magnetic 
field increases, while the $U/I$ wing signals first increase and then 
decrease. Also in this case, when the same calculations are carried out 
neglecting the $\rho_V$ terms, only the line-center signals are modified 
with respect to the zero-field case.  

In terms of linear polarization degree, $p_L = \frac{\sqrt{Q^2 +U^2}}{I}$, and linear polarization angle, 
the considered vertical and horizontal field cases of increasing intensity produce in the near 
wings (where $Q/I$ reaches its largest negative signal) a monotonic decrease of $p_L$ and a 
monotonic rotation of the plane of linear polarization.

%%%%%%%%%%%%%%%%%%%%%%%%%%%%%%%%%%%%%%%%%%%%%%%%%%%%%%%%%%%%
\subsection{Disk center geometry (LOS with $\mu=1$)}
For the case of a vertical magnetic field we only have $V/I$ in the disk center geometry considered in this section.
The left panel of Fig.~5 shows the $V/I$ profiles of the emergent Mg~{\sc ii} 
$k$ line radiation, for a vertical field with B=50~G. 
The red solid curve shows the full solution, while the dashed curve has been 
obtained assuming that the only non-zero component of the radiation 
field tensor is $J^0_0$ (i.e., the familiar mean intensity). Interestingly, the 
main $V/I$ peaks close to the line-center basically coincide in the two cases, 
while the secondary lobes in the near wings of the line are significantly 
weaker when the symmetry properties of the incident radiation field (quantified 
by the $J^1_Q$ and $J^2_Q$ 
components) are not taken 
into account. This demonstrates that the wing lobes of the $V/I$ profiles are partly 
influenced by the symmetry properties of the radiation field. 
In spite of this fact, we point out that a good estimation of the longitudinal 
component of the magnetic field can still be obtained using the familiar 
weak-field magnetograph formula for Stokes $V$.

The right panel of Fig.~5 shows the $Q/I$ profiles produced by horizontal 
fields of increasing strength.  
Obviously, in the zero-field case $Q/I=0$. 
The axial symmetry around the local vertical direction is broken in the 
presence of a horizontal magnetic field. 
As a result, the Hanle effect creates linear polarization at the very line 
center, with a $Q/I$ amplitude that increases with the magnetic strength till 
the Hanle saturation regime is reached. All this is well known. 
What is new and interesting are the $Q/I$ wing signals.
They are still due to magneto-optical effects produced by $\rho_V$, though in 
an indirect way. Indeed, $\rho_V$ is zero for the radiation propagating in the $\mu=1$ direction
(i.e. perpendicularly to the magnetic field). However the emissivities $\epsilon_Q$ and $\epsilon_U$
 appearing in Eqs.~(3) and (4) depend (through the redistribution matrix) on the pumping radiation 
coming from {\em all} directions, whose wings are sensitive to magneto-optical effects. 
Due to the coherency of scattering, such sensitivity is transferred to the radiation scattered at 
$\mu=1$. 

%%%%%%%%
\begin{figure}[t]
\begin{center}
\includegraphics[scale=0.28]{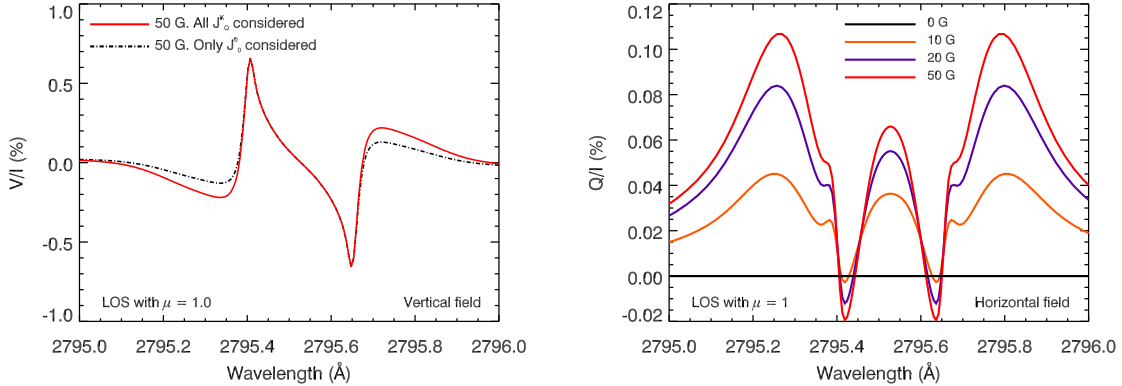}
\caption{Left panel: the $V/I$ profiles produced by the Zeeman effect of a 
vertical field with $B=50$ G, with the red solid curve showing the full solution 
and the black dashed-dotted one that obtained assuming that the familiar mean intensity is 
the only non-zero component of the radiation field tensor. 
Right panel: the $Q/I$ profiles of the Mg {\sc ii} $k$ line calculated in the 
FAL-C semi-empirical model, in the absence (black lines) and in the 
presence (colored curves) of a horizontal magnetic field. 
The $Q/I$ results for 100 G and 200 G (not shown) are similar to the red curves. All the
 results refer to a LOS with $\mu = 0.1$. The reference direction for positive Stokes $Q$ is the perpendicular 
 to the magnetic field.}
\label{Figure-5}
\end{center}
\end{figure} 
%%%%%%%%%

\section{Concluding comments}

We have investigated the magnetic sensitivity of the Mg {\sc ii} $k$ line
at 2795.5 \AA\ due to the joint action of Hanle, Zeeman and magneto-optical
effects. To this end, we applied a new RT code for a two-level atom 
\citep[see][]{Alsina+16}. This atomic model is suitable for
investigating the polarization of the Mg {\sc ii} $k$ line between the $k_{1{\rm V}}$ and
$k_{1{\rm R}}$ intensity peaks (see Figs. 3 or 4), where the effects of
$J$-state interference are not very significant.
This is arguably the most interesting spectral region of the Mg {\sc ii}
$h$ \& $k$ line system because in this wavelength interval the line
opacity is relatively large, so that it encodes information on the physical
properties of the solar chromosphere \citep[see figure 1 in][]{BelluzziTrujillo12}.
Moreover, only the Mg {\sc ii} $k$ line is sensitive to the Hanle effect,
and in such spectral region the sensitivity of the line scattering polarization
to the thermal structure of the solar atmosphere is not as important as in the
far wings of the Mg {\sc ii} $h$ \& $k$ lines, which can be useful to
facilitate the inference of the magnetic field.

Via the Hanle effect the line-center amplitudes of the $Q/I$ and $U/I$
profiles of the Mg {\sc ii} $k$ line are sensitive to magnetic fields with strengths
between 5 and 50 G, approximately. Moreover, the Zeeman effect produces
conspicuous circular polarization signals, with two main lobes just around
the line center surrounded by two wing lobes whose weaker amplitudes are
sensitive to the symmetry properties of the pumping radiation field. Nevertheless,
the familiar magnetograph formula can be used to estimate the longitudinal
field component.

The most novel result is that the magneto-optical $\rho_V$ terms of the
Stokes $Q$ and $U$ transfer equations tend to depolarize the wings of the $Q/I$
scattering polarization pattern and create signals in the wings of $U/I$.
The sensitivity of these $Q/I$ and $U/I$ wing signals to both weak
(${\sim}\,5$ G) and stronger magnetic fields expands the scientific
interest of the Mg {\sc ii} $k$ line for probing the solar chromosphere in quiet and
active regions of the Sun.

Since our two-level atom approach cannot account for the effects of
$J$-state interference, we cannot provide quantitative information outside the
spectral region between the $k_{1{\rm V}}$ and $k_{1{\rm R}}$ peaks, but it
is clear that the same magneto-optical effects are expected to affect
also the spectral region between the $h$ and $k$ lines, as well as their far
wings. A correct modeling of such spectral regions requires the application of a two-term or multi-term 
atom approach in order to take into account the impact of $J$-state interference.

\acknowledgements
The authors are grateful to T. del Pino Alem\'an, R. Casini (both HAO), and R. Manso Sainz (MPS), who have been working on a similar problem applying a recently-developed multi-term formalism, for stimulating scientific conversations. 
Financial support by the Spanish Ministry of Economy and Competitiveness through projects \mbox{AYA2014-60476-P} and \mbox{AYA2014-55078-P} is gratefully acknowledged. E. Alsina Ballester wishes to acknowledge the Fundaci\'on La Caixa for financing his Ph.D. grant. L. Belluzzi gratefully acknowledges financial support from SERI, Canton Ticino, the city of Locarno, local municipalities, Dacc\`o foundation, and the Swiss SNF through grant 200021-163405


\begin{thebibliography}{}
\expandafter\ifx\csname natexlab\endcsname\relax\def\natexlab#1{#1}\fi

\bibitem[{{Alsina Ballester} {et~al.}(2016){Alsina Ballester}, {Belluzzi}, \& {Trujillo Bueno}}]{Alsina+16}
{Alsina Ballester}, E., {Belluzzi}, L., \& {Trujillo Bueno}, J. 2016, \apj, in press

\bibitem[{{Belluzzi} \& {Trujillo Bueno}(2011)}]{BelluzziTrujillo11}
{Belluzzi}, L., \& {Trujillo Bueno}, J. 2011, \apj, 743, 3

\bibitem[{{Belluzzi} \& {Trujillo Bueno}(2012)}]{BelluzziTrujillo12}
---. 2012, \apjl, 750, L11

\bibitem[{{Belluzzi} \& {Trujillo Bueno}(2014)}]{BelluzziTrujillo14}
---. 2014, \aap, 564, A16

\bibitem[{{Belluzzi} {et~al.}(2012){Belluzzi}, {Trujillo Bueno}, \& {{\v S}t{\v
  e}p{\'a}n}}]{Belluzzi+12}
{Belluzzi}, L., {Trujillo Bueno}, J., \& {{\v S}t{\v e}p{\'a}n}, J. 2012,
  \apjl, 755, L2
  
\bibitem[{{Bommier} (1997)}]{Bommier97} 
{Bommier}, V. 1997, \aap, 328, 726
 
\bibitem[{{Fontenla} {et~al.}(1993){Fontenla}, {Avrett}, \&
  {Loeser}}]{Fontenla+93}
{Fontenla}, J.~M., {Avrett}, E.~H., \& {Loeser}, R. 1993, \apj, 406, 319

\bibitem[{{Kano} {et~al.}(2016){Kano}, {Trujillo Bueno}, {Winebarger}, \& {et
  al.}}]{Kano+16}
{Kano}, R., {Trujillo Bueno}, J., {Winebarger}, A., {et al.} 2016, in Solar Polarization 8, 
ASP Conf. Series, Vol. in press

\bibitem[{{Landi Degl'Innocenti} \& {Landolfi}(2004)}]{LL04}
{Landi Degl'Innocenti}, E., \& {Landolfi}, M. 2004, Polarization in Spectral
  Lines (Klumer Academic Publishers)

\bibitem[{{Stenflo}(1994)}]{Stenflo94}
{Stenflo}, J., ed. 1994, Astrophysics and Space Science Library, Vol. 189,
  {Solar Magnetic Fields: Polarized Radiation Diagnostics}

\bibitem[{{Trujillo Bueno}(2014)}]{Trujillo14}
{Trujillo Bueno}, J. 2014, in Astronomical Society of the Pacific Conference
  Series, Vol. 489, Solar Polarization 7, ed. K.~N. {Nagendra}, J.~O.
  {Stenflo}, Q.~{Qu}, \& M.~{Sampoorna}, 137
  
\bibitem[{{Trujillo Bueno} {et~al.}(2011){Trujillo Bueno}, {{\v S}t{\v
  e}p{\'a}n}, \& {Casini}}]{Trujillo+11}
{Trujillo Bueno}, J., {{\v S}t{\v e}p{\'a}n}, J., \& {Casini}, R. 2011, \apjl,
  738, L11  

\bibitem[{{Trujillo Bueno} {et~al.}(2012){Trujillo Bueno}, {{\v S}t{\v
  e}p{\'a}n}, \& {Belluzzi}}]{Trujillo+12}
{Trujillo Bueno}, J., {{\v S}t{\v e}p{\'a}n}, J., \& {Belluzzi}, L. 2012,
  \apjl, 746, L9

\bibitem[{{{\v S}t{\v e}p{\'a}n} {et~al.}(2012){{\v S}t{\v e}p{\'a}n},
  {Trujillo Bueno}, {Carlsson}, \& {Leenaarts}}]{Stepan+12}
{{\v S}t{\v e}p{\'a}n}, J., {Trujillo Bueno}, J., {Carlsson}, M., \&
  {Leenaarts}, J. 2012, \apjl, 758, L43

\end{thebibliography}
\end{document}